\documentclass[twocolumn,traditabstract,longauth]{aa}
\usepackage{graphicx}

\usepackage[T1]{fontenc}
\usepackage[a4paper]{geometry}
\makeatletter
\providecommand{\tabularnewline}{\\}
\usepackage{pdflscape}
\makeatother

\begin{document}

\title{H$_2$O emission in high-$z$ ultra-luminous infrared galaxies \thanks{{\it Herschel} (Pilbratt et al. (2010) is 
an ESA space observatory with science instruments provided by European-led Principal Investigator consortia 
and with important participation from NASA.} 
}

\author{A. Omont\inst{1,2}
\and Chentao Yang\inst{3,1,2,4}
\and P. Cox\inst{5}
\and R. Neri\inst{5}
\and A. Beelen\inst{6}
\and S. Bussmann\inst{7}
\and R. Gavazzi\inst{1,2}
\and P. van der Werf\inst{8}
\and D. Riechers\inst{9,28}
\and D. Downes\inst{5}
\and M. Krips\inst{5}
\and S. Dye\inst{10}
\and R. Ivison\inst{11}
\and J.D. Vieira\inst{28}
\and A. Wei\ss\inst{20}
\and J.E. Aguirre\inst{12}
\and M. Baes\inst{13}
\and A.J. Baker\inst{14}
\and F. Bertoldi\inst{15}
\and A. Cooray\inst{18}
\and H. Dannerbauer\inst{16}
\and G. De Zotti\inst{21}
\and S.A. Eales\inst{17}
\and H. Fu\inst{18}
\and Y. Gao\inst{3}
\and M. Gu\'elin\inst{5}
\and A.I. Harris\inst{19}
\and M. Jarvis\inst{24,34}
\and M. Lehnert\inst{1,2,31}
\and L. Leeuw\inst{36}
\and R. Lupu\inst{12}
\and K. Menten\inst{20}
\and M.J.~Micha{\l}owski\inst{13,11}
\and M. Negrello\inst{21}
\and S. Serjeant\inst{22}
\and P. Temi\inst{23}
\and R. Auld\inst{17} 
\and A. Dariush\inst{25,17}
\and L. Dunne\inst{26,27} 
\and J. Fritz\inst{13}
\and R. Hopwood\inst{35}
\and C. Hoyos\inst{27}
\and E. Ibar\inst{32,33}
\and S. Maddox\inst{26,27}
\and M.W.L. Smith\inst{17}
\and E. Valiante\inst{17}
\and J. Bock \inst{28,29}
\and C.M. Bradford \inst{28,29}
\and J. Glenn \inst{30}
\and K.S. Scott \inst{12}
}
\institute{UPMC Univ Paris 06, UMR7095, Institut d'Astrophysique de Paris, 
F-75014, Paris, France
\and  CNRS, UMR7095, Institut d'Astrophysique de Paris, F-75014, Paris, France 
\and Purple Mountain Observatory, Nanjing, China
\and Department of Astronomy, Beijing Normal University, Beijing, China
\and  Institut de Radioastronomie Millim{\'e}trique (IRAM), 300 rue de la Piscine, 38406 Saint-Martin d'H{\`e}res, France
\and  Univ Paris-Sud and CNRS, Institut d'Astrophysique Spatiale, UMR8617, Orsay, F-91405, France
\and Harvard-Smithsonian Center for Astrophysics, 60 Garden Street, Cambridge, MA 02138, USA
\and  Leiden Observatory, Leiden University, Post Office Box 9513, NL - 2300 RA Leiden, The Netherlands 
\and Department of Astronomy, Space Science Building, Cornell University, Ithaca, NY, 14853-6801, USA;  
\and Centre for Astronomy \& Particle Theory, University Park, Nottingham, NG7 2RD, UK
\and  Institute for Astronomy, University of Edinburgh, Royal Observatory, Bford Hill, Edinburgh EH9 3HJ
\and  Department of Physics and Astronomy, University of Pennsylvania, Philadephia, PA 19104, US
\and  Sterrenkundig Observatorium, Universiteit Gent, Krijgslaan 281-S9, B-9000 Gent, Belgium 
\and  Department of Physics and Astronomy, Rutgers, The State University of New Jersey, Piscataway, NJ 08854-8019, USA
\and  Argelander Institut fur Astronomie, Universitat Bonn, Auf dem Hugel 71, 53121 Bonn, Germany
\and Universitat Wien, Institut fur Astronomie, Turkenschanz-strasse 17, 1180 Wien, Austria
\and  School of Physics and Astronomy, Cardiff University, Queens Buildings, The Parade 5, Cardiff, CF24 3AA, UK
\and  Department of Physics and Astronomy, University of California, Irvine, CA 92697, USA
\and Department of Astronomy, University of Maryland, College Park, MD 20742, USA
\and  Max-Plank-Institut fur Radioastronomie, Auf dem Hugel 69, 53121 Bonn, Germany
\and  INAF-Osservatorio Astronomico di Padova, Vicolo Osservatorio 5, I-35122 Padova, Italy; and SISSA, Via Bonomea 265, I-34136 Trieste, Italy
\and  Department of Physics and Astronomy, The Open University, Walton Hall, Milton Keynes, MK7 6AA, UK
\and   Astrophysics Branch, NASA Ames Research Center, MS 245-6, Moffett Field, CA 94035, USA
\and Astrophysics, Department of Physics, Keble Road, Oxford OX1 3RH, UK
\and  School of Astronomy, Institute for Research in Fundamental Sciences (IPM),PO Box 19395-5746, Tehran, Iran 
\and Department of Physics and Astronomy, University of Canterbury, Private Bag 4800, Christchurch, 8140, NZ
\and School of Physics and Astronomy, University of Nottingham, University Park, Nottingham NG7 2RD, UK
\and  California Institute of Technology, 1200 E. California Blvd, Pasadena, CA 91125, USA
\and  Jet Propulsion Laboratory, Pasadena, CA 91109, USA
\and  University of Colorado, CASA 389-UCB, Boulder, CO 80303, USA
\and  GEPI, Observatoire de Paris, CNRS, Universit\'e Paris Diderot, 5 Place Jules Janssen, 92190 Meudon, France
\and UK Astronomy Technology Centre, The Royal Observatory, Blackford Hill, Edinburgh, EH9 3HJ, UK
\and Universidad Catolica de Chile, Departamento de Astronomia y Astrofisica, Casilla 306, Santiago 22, Chile
\and Department of Physics, University of the Western Cape, Private Bag X17, Bellville 7535, South Africa
\and Astrophysics Group, Imperial College, Blackett Laboratory, Prince Consort Road, London, SW7 2AZ, UK
\and College of Graduate Studies, University of South Africa, P. O. Box 392, Unisa, 003, South Africa 
}



\abstract{Using the IRAM Plateau de Bure interferometer (PdBI), we report the detection of water vapor in
six new lensed ultra-luminous starburst galaxies at high redshift, discovered in the {\it Herschel} 
Astrophysical Terahertz Large Area Survey (H-ATLAS). 
The sources are detected either in the
2$_{02}$--1$_{11}$  or 2$_{11}$--2$_{02}$ H$_2$O emission lines with 
integrated line fluxes ranging 
from 1.8 to 14~Jy~km~s$^{-1}$. The corresponding apparent luminosities are $\rm \mu L_{H_2O} \sim 3-12\,x\,10^8 \, L_{\sun}$,
where $\mu$ is the lensing magnification factor ($\rm 3 < \mu < 12$). These results confirm that H$_2$O lines are 
among the strongest molecular lines in high-$z$ ultra-luminous starburst galaxies, with intensities almost comparable 
to those of the high-J CO lines, and similar profiles and line widths ($\sim$200-900~km~s$^{-1}$). 
With the current sensitivity of the PdBI, the water lines can 
therefore easily be detected in high-$z$ lensed galaxies (with F(500$\mu$m)\,$>$\,100\,mJy) discovered in the 
{\it Herschel} surveys. 
Correcting the luminosities for amplification, using existing lensing models,  
L$_{\rm H_2O}$  is found to have a strong dependence on the 
infrared luminosity, varying as $\sim$L$_{\rm IR}^{1.2}$. 
This relation, which needs to be confirmed with better statistics,
may indicate a role of radiative (infrared) excitation of the H$_2$O lines,
and implies that high-$z$ galaxies with 
L$_{\rm IR}$\,$\gtrsim$\,10$^{13}$\,$\rm L_{\sun}$ tend to be very strong emitters in water vapor, that have no 
equivalent in the local universe.}

\keywords{Galaxies: high-redshift -- Galaxies: starburst -- Galaxies:
active -- Infrared: galaxies -- Submillimeter: galaxies -- Radio lines:
galaxies}

\maketitle

\section{Introduction}

Water plays an important role in the warm dense interstellar medium of galaxies. First, after CO, H$_2$O is the most abundant oxygen-bearing molecule, and, second, it can be an important 
coolant of the warm gas. Due to the Earth's atmosphere, bulk gas-phase water can only be detected 
from space or from the ground toward distant objects with redshifts that move the H$_2$O lines 
into atmospheric windows. 

The study of water emission lines in nearby galaxies has recently made significant progress thanks to 
the availability of their infrared/submillimeter spectra using the spectrometer mode of the {\it Herschel} Spectral and 
Photometric Imaging REceiver (SPIRE, Griffin et al.\ 2010) and the Photodetecting Array 
Camera and Spectrometer (PACS, Poglitsch et al.\ 2010). The spectra of local  
ultra-luminous infrared galaxies (ULIRGs) and composite AGN/starbursts such as Mrk\,231
(van der Werf et al.\ 2010), Arp\,220 (Rangwala et al.\ 2011; Gonz\'alez-Alfonso et al.\ 2012), and 
NGC\,4418 (Gonz\'alez-Alfonso et al.\ 2012) reveal a wealth of water lines and the presence of 
associated molecules such as OH$^+$, $\rm H_2O^+$ and isotopologues. Together with high-J CO lines, 
these lines provide an important diagnostic of the warm dense cores of nearby ULIRGs. 

In the cases of Mrk\,231 and Arp\,220, water emission lines up to energy levels of E$_{\rm up}$/k = 642\,K 
are detected with strong line fluxes that reach 25\%--75\% of the neighboring CO emission 
line fluxes. Spectral surveys made with {\it Herschel} show that low-$z$ ULIRGs always exhibit bright H$_2$O 
lines, whereas, only one third of the sample of luminous infrared galaxies (LIRGs) 
displays luminous $\rm H_2O$ emission lines (van der Werf et al.\ in preparation [prep.]), indicating that the strength of the
water lines and the infrared luminosity, $\rm L_{IR}$, must be related. 
The analysis of the H$_2$O emission lines in Mrk\,231 shows that the excitation of the water molecules
results from a combination of collisions and infrared excitation through far-infrared lines in warm dense gas ($\rm \ge \, 100 \, K$, 
$\rm \ge 10^5 \, cm^{-3}$). Moreover, the far-infrared radiation field dominates the excitation of the 
high levels and their emission lines (Gonz\'alez-Alfonso et al.\ 2010). Preliminary evidence from the comparison of the 
spectra of  Arp\,220,  Mrk\,231 and NGC\,4418 shows that the properties of the 
water emission lines in their nuclear regions vary as a function of chemistry, nucleosynthesis, and 
inner motions (outflow/infall) - see Rangwala et al.\ (2011) and Gonz\'alez-Alfonso et al.\ (2012).  

First detections of H$_2$O megamasers at high redshift were reported in objects at $z=0.66$ 
and 2.64 by Barvainis \& Antonucci (2005) 
and Impellizzeri et al. (2008), while further searches have failed since then (McKean et al.\ 2011). Following initial 
attempts to detect $\rm H_2O$ rotational emission lines from high-$z$ galaxies (Wagg et al.\ 2006; Riechers et 
al.\ 2006, 2009) and tentative detections in IRAS F10214+4724 at $z = 2.23$ (Casoli et al.\ 1994) and in the Cloverleaf at $z = 2.56$ (Bradford et al.\ 2009), a series of robust detections of non-maser $\rm H_2O$ emission lines were recently reported in high-$z$ sources. 
The sources detected in $\rm H_2O$ emission lines include a strongly lensed galaxy, HATLAS~J090302.9$-$014127 
(SDP.17b) at $z=2.3$ (Omont et al.\ 2011), which was uncovered in the {\it Herschel} Astrophysical 
Terahertz Large Area Survey (H-ATLAS) (Eales et al.\ 2010); the gravitationally 
lensed quasar APM\,08279+5255 at $z=3.9$ (van der Werf et al.\ 2011; Bradford et al.\ 2011; Lis et al.\ 2011); 
HLSJ091828.6+514223, a lensed submillimeter galaxy at $z=5.2$ in the field of Abell 773 
(Combes et al.\ 2012), which was found in the {\it Herschel} Lensing Survey (Egami et al.\ 2010), IRAS F10214+4724 (Riechers et al., in prep.). 
For most of these sources, only one water emission line was reported, with the exception
of APM\,08279+5255, where a total of at least five emission lines with  $\rm E_u/k = 101$ to  454\,K  
were detected (van der Werf et al.\ 2011; Lis et al.\ 2011; Bradford et al.\ 2011). These results underline the unique and 
powerful diagnostic power of H$_2$O emission lines, which give better insight into local conditions in distant
galaxies than may be obtained by other means. As in Mrk\,231, they reveal the presence of extended,  
warm and dense gas located in the infrared-opaque regions of the galactic cores.

This new window in the exploration of high-$z$ sources is based on the combined availability of new 
instrumentation with improved sensitivities at key facilities and the increasing number of 
gravitationally lensed sources discovered in the {\it Herschel} and South Pole Telescope (SPT) 
cosmological surveys 
(see, e.g., Negrello et al. 2010; Vieira et al.\ 2010). Following the results reported by Omont et al. (2011), we 
present here a new study of water emission in six high-$z$ lensed ULIRGs, which were selected 
from the H-ATLAS survey. The data, which were obtained using the IRAM Plateau de Bure Interferometer 
(PdBI), clearly show water emission lines in all the sources. 
Based on these results, we derive a clear relation between the $\rm H_2O$ and the infrared 
luminosities in high-$z$ ULIRGs.   

Throughout this paper, we adopt a cosmology with $H_{0}=71\,{\rm km\,s^{-1}\,Mpc^{-1}}$,
$\Omega_{M}=0.27$, $\Omega_{\Lambda}=0.73$ (Spergel et al.\ 2003).

\begin{figure*}
\begin{center}
{\small \includegraphics[scale=1.0]{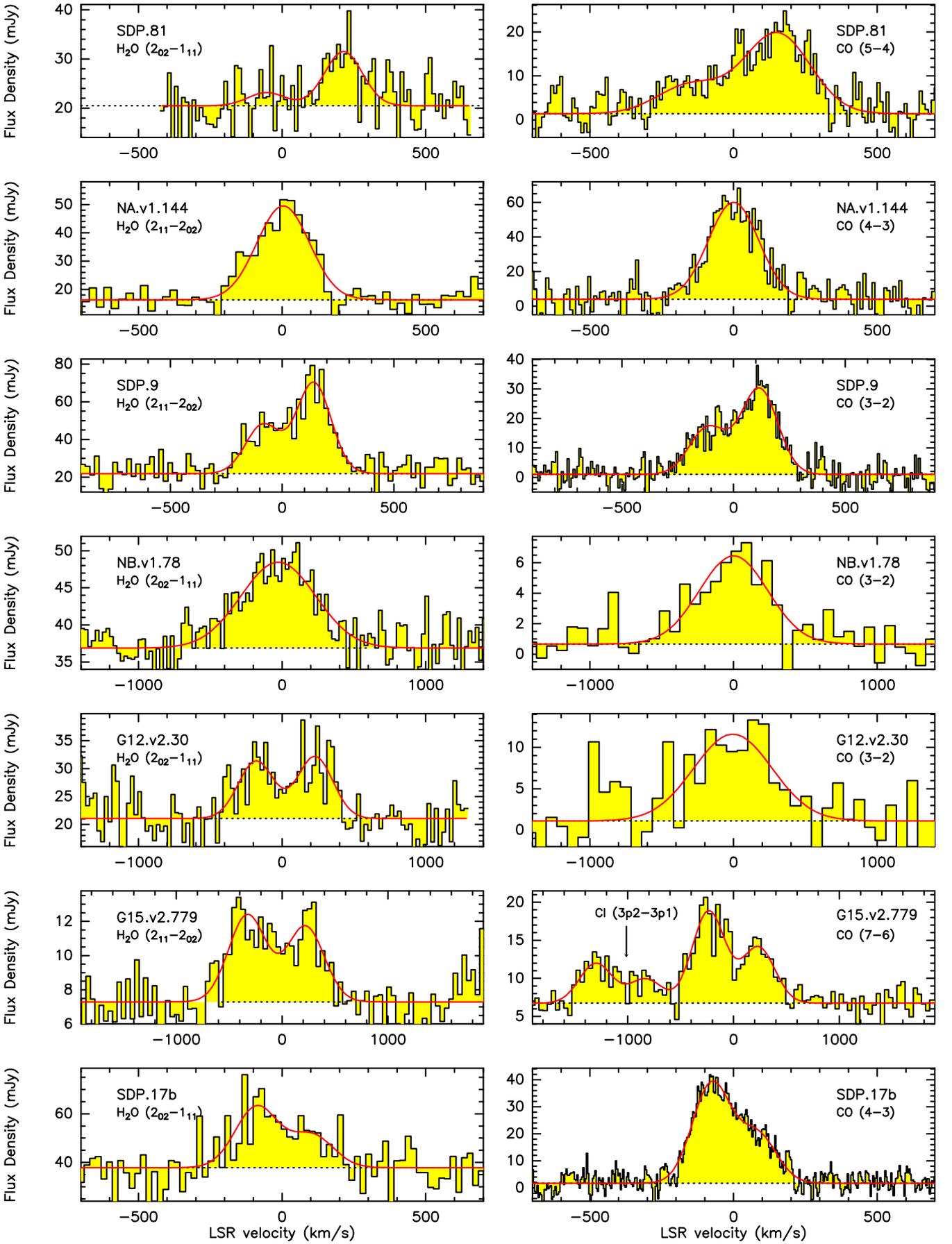}}
\caption{Spectra of the $\rm H_2O$ and CO emission lines observed at PdBI toward the seven lensed high-redshift 
ultra-luminous galaxies discussed in this paper. 
The $\rm H_2O$ emission lines are displayed in the left column (showing the 
$(2_{11}-2_{02})$ or the $(2_{02}-1_{11})$ transition; the transitions are identified in each box. The CO spectra are displayed in the right column
(showing 3--2, 4--3, 5--4 and 7--6 rotational transitions - for G15.v2.779, the [C{\small I}] emission line
is included in the band). The transitions are indentified in each box.  
All spectra include the continuum, and the $\rm H_2O$ frequencies corresponding to the zero velocity are given in Table 4.  
The red lines trace the Gaussian fits to the $\rm H_2O$ and CO line profiles and the fit to the continuum emission level. 
In all cases, the match between the $\rm H_2O$ and CO spectra is excellent both in redshift and in the 
details of the line profile - see text.} 
\par\end{center}
\end{figure*}

\section{Sample selection and observations}

The detection of strong $\rm H_2O$ emission in the lensed H-ATLAS galaxy SDP.17b by Omont et al. (2011) suggested that 
the H-ATLAS survey offered a unique opportunity to select an homogeneous sample of bright lensed galaxies spanning 
a broad range in luminosity and redshift to further study the properties of water emission in the early universe. 
We selected from the H-ATLAS catalog, sources with strong flux densities (with F$_{500\mu m}$\,$>$\,150~mJy) 
that were well characterized, i.e. had CO redshift measurements, additional sub/millimeter imaging 
and available  deflector identification (see, e.g., Negrello et al. 2010; Bussmann et al.\ in prep.). 
In the selection, 
we also somewhat preferred sources at $z>3$ to shift the H$_2$O lines in the low frequency bands that are easier to observe.
Finally, we chose sources spanning a wide range in intrinsic infrared luminosities from $\sim \,5 \times 10^{12}$ to 
$\rm \sim \, 2 \times 10^{13} \, L_{\sun}$. 

In this paper, we report the first results of this survey and describe the properties of the $\rm H_2O$ emission
for six new sources. The sample includes (see Table 1 for IAU names of the sources) i) sources reported 
in the initial H-ATLAS Science Demonstration Phase paper by Negrello et al. (2010) - SDP.9 and SDP.81, 
together with SDP.17b, which was discussed in Omont et al. (2011); 
ii) two other well-studied sources from the H-ATLAS equatorial fields - G12.v2.30 (Fu et al. 2012) and 
G15.v2.779 (Cox et al. 2011; Bussmann et al. 2012); 
iii) two sources from the H-ATLAS NGP field, NA.v1.144 and NB.v1.78, for which CO observations 
(Harris et al.\ 2012; Riechers et al.\ in prep.; and this paper) and 
submillimeter imaging (Bussmann et al. in prep.) are available. Table~1 provides details on the $\rm H_2O$ observations 
of the seven H-ATLAS lensed galaxies, Table~2 lists their submillimeter and infrared properties together 
with the estimated amplification factors (see Sects.~3.2 and 4.1), and Table~3 gives their CO properties available 
from recent measurements using the PdBI. 

The redshifts of the seven selected galaxies range from 1.58 to 4.24, with the majority in the range $2.0<z<3.3$. 
In the sample, G15.v2.779 is the only source at $z>4$ and, for the time being, the highest redshift lensed galaxy  
spectroscoptically confirmed in the H-ATLAS survey. For each source, at least one of the two strongest low-excitation H$_2$O lines is 
in an atmospheric window observable with the PdBI, either H$_2$O(2$_{11}$--2$_{02}$) with $\nu _{\rm rest}$=752.033\,GHz and 
E$_{\rm up}$\,=\,137\,K, or H$_2$O(2$_{02}$--1$_{11})$ with $\nu _{\rm rest}$=987.927\,GHz and E$_{\rm up}$\,=\,101\,K. 
As both lines have comparable intensities in Arp\,220 and Mrk\,231 (see Sec.~4), in the rare cases where both 
lines are observable, we chose the line at the lower frequency that is easier to observe, except in the case 
of SDP.17b (Omont et al.\ 2011), where the line  H$_2$O(2$_{02}$--1$_{11})$ was selected to confirm the tentative 
detection that was reported by Lupu et al. (2012).
Note that these two observed lines are both para-H$_2$O and they are adjacent lines in the H$_2$O level diagram.

The H$_2$O observations were conducted in the compact D-configuration from December 2011 to March 2012 in conditions of good atmospheric
phase stability (seeing of $1''$) and reasonable transparency ($\rm PWV \le 1 \, mm$). Except for SDP.81, which was observed
with five antennas, all other sources were observed with the six antennas of the PdBI array. The total on-source
integration ranges from $\sim$1 to 2 hours (Table~1). We used Bands 2, 3 and 4, which cover the frequency ranges 
129--174~GHz, 201--267~GHz and 277--371~GHz, and the band centers were tuned to the redshifted frequency 
of the selected $\rm H_2O$ emission line using the redshifts estimated from CO observations (Table 3). The correlator (WideX) 
provided a contiguous frequency coverage of 3.6~GHz in dual polarization with a fixed channel spacing of 1.95~MHz, 
allowing us to detect the continuum as well as any additional emission lines, if present. 
In the compact configuration, the baselines extend from 24 to 179\,m, resulting in synthesized beams of 
$\sim 1.5''$ x $1.0''$ to $\sim 4''$ x $3''$ (Table 1). Only SDP.81, G12.v2.30 and NB.v1.78  
are resolved at this angular resolution (Fig.\ 2). 

During the observations, the phase and bandpass were calibrated by measuring standards sources that are regularly 
monitored at the IRAM PdBI, including 3C84, 3C279, MWC349,  CRL618, and 0851+202. The accuracy of the flux calibration 
is estimated to range from $\sim$10\% at 2~mm to $\sim$20\% at 0.8~mm.

To complement the $\rm H_2O$ data, we include in this paper new results on the CO emission for the seven sources. For five 
of them, the CO data were obtained in 
2011 and 2012 using the PdBI in the A-configuration. Those data
were acquired in the frame of a survey of lensed ULIRGs to map their CO and dust continuum emission at high-angular resolution 
(Cox, Ivison et al.\ in prep.) and a full description of the results will be given in that paper. 
For two of the sources, NB.v1.78 and G12.v2.30, the CO data were obtained in August 2012 in the D-configuration 
with five antennas. The observations were made in good weather conditions and more details are provided in Table~3. 
In this paper, we only present the global CO spectra (i.e., integrated over the source's extent) with the goal to 
compare the characteristics of the $\rm H_2O$ and CO emission line profiles. A detailed discussion of the morphology, 
the dynamics, and the lensing of these sources will be given in Cox, Ivison et al.\ (in prep.).  

\begin{figure*}
\begin{center}
{\small \includegraphics[scale=1.0]{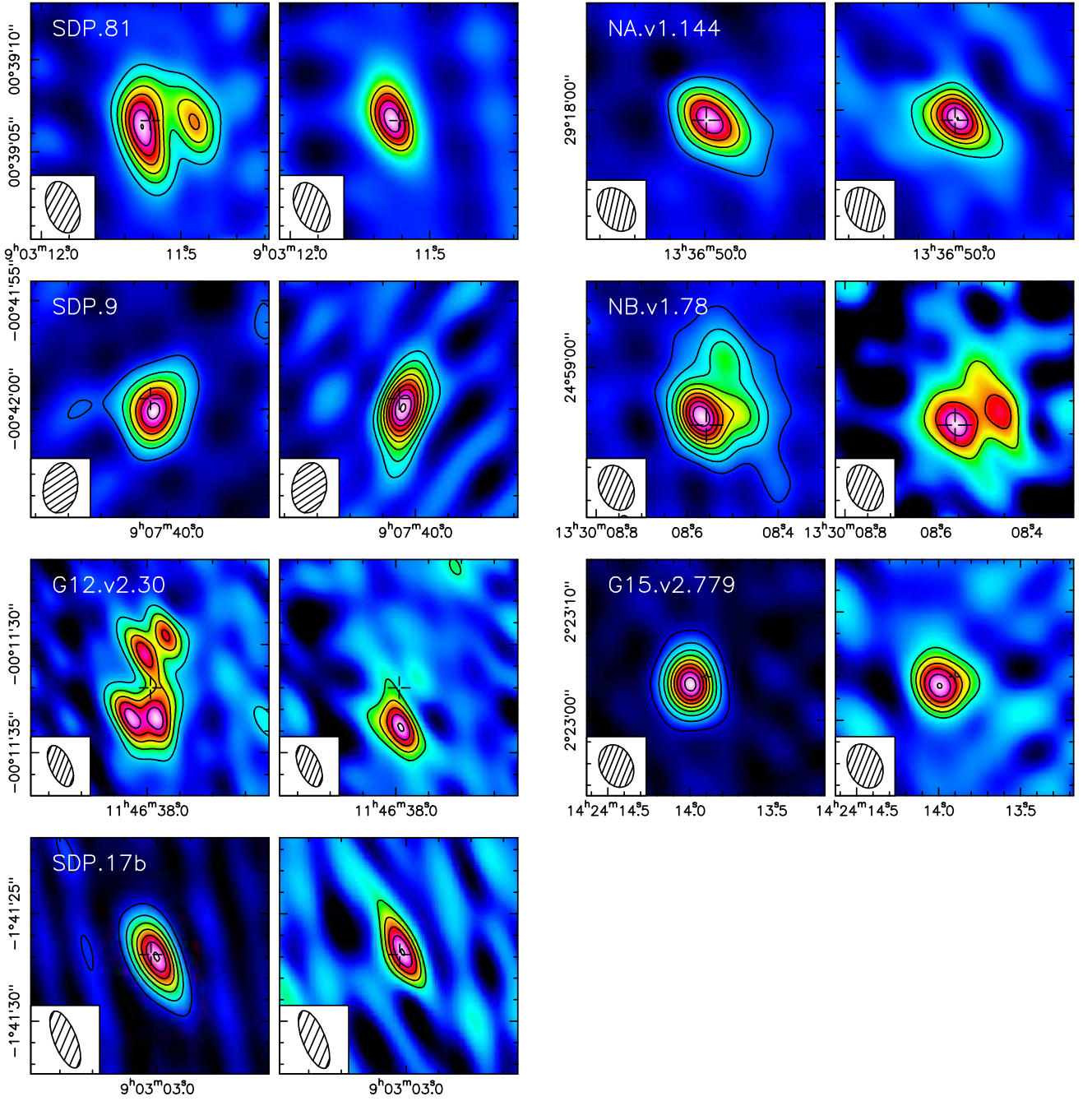}}
\caption{Images of the submillimeter continuum (left panel) and $\rm H_2O$ line emission (right panel - see Fig.\ 1 and Table 4 for the identification of the transitions) for the
         seven high-$z$ lensed ultra-luminous galaxies studied in this paper (sources are identified in the
         left panels). Synthesized beams are shown in the lower left corner of each map. The map centers 
         correspond to the coordinates given in Table~1 and are indicated by a cross.  For the continuum
         maps, the contours start from 3$\sigma$ in steps of 3$\sigma$ for SDP.81 and G12.v2.30, and in steps
         of 6$\sigma$ for the other sources; for the $\rm H_2O$ line emission maps, the contours start 
         from 3$\sigma$ in steps of 1$\sigma$ (for SDP.81) and 2$\sigma$
         for the other sources. For each source, the 1$\sigma$ continuum (mJy/beam) and
         emission line (Jy~km~s$^{-1}$/beam) noise levels are as follows: SDP.81 (0.60 / 0.18),
         NA.v1.144 (0.35 / 0.31),
         SDP.9 (0.50 / 0.59), NB.v1.78 (0.28 / 0.30), G12.v2.30 (0.28 / 0.28), G15.v2.779 (0.11 / 0.24), and
         SDP.17b (0.73 / 0.56). Four sources remain unresolved at the current angular resolution,
         whereas SDP.81, NB.v1.78, and G12.v2.30 display complex lensing morphologies both in the continuum and line emission
         that are revealed even at the present low angular resolution. Detailed comments on the sources are given
         in the text (Section~3).}
\par\end{center}
\end{figure*}

\begin{table*}
\centering
\caption{Observation Log}
\begin{center}
{\scriptsize }%
\begin{tabular}{cccccccc}
\multicolumn{8}{c}{}\tabularnewline
\hline
\hline
{\scriptsize IAU Name} & { ID} & { $\mathrm{RA^{a}}$} & { $\mathrm{Dec{}^{a}}$} & { Date} & { $\mathrm{Frequency^{b}}$} & { Beam($''$)} & { $\mathrm{t_{on}(h)^{c}}$}\tabularnewline
\hline
{\scriptsize HATLAS J090311.6$+$003906} & {\scriptsize SDP.81} & {\scriptsize 09:03:11.61} & {\scriptsize $+$00:39:06.7} & {\scriptsize 2011 Dec} & {\scriptsize 244.890} & {\scriptsize 2.8$\times$1.8} & {\scriptsize 1.2}\tabularnewline
{\scriptsize HATLAS J133649.9$+$291801} & {\scriptsize NA.v1.144} & {\scriptsize 13:36:50.00} & {\scriptsize $+$29:17:59.6} & {\scriptsize 2012 Mar} & {\scriptsize 235.330} & {\scriptsize 1.6$\times$1.1} & {\scriptsize 0.8}\tabularnewline
{\scriptsize HATLAS J090740.0$-$004200} & {\scriptsize SDP.9} & {\scriptsize 09:07:40.05} & {\scriptsize $-$00:41:59.5} & {\scriptsize 2011 Dec} & {\scriptsize 292.550} & {\scriptsize 3.3$\times$1.6} & {\scriptsize 1.4}\tabularnewline
{\scriptsize HATLAS J133008.4$+$245900} & {\scriptsize NB.v1.78} & {\scriptsize 13:30:08.56} & {\scriptsize $+$24:58:58.3} & {\scriptsize 2012 Jan \& Mar} & {\scriptsize 240.802} & {\scriptsize 1.4$\times$1.0} & {\scriptsize 2.3}\tabularnewline
{\scriptsize HATLAS J114637.9$-$001132} & {\scriptsize G12.v2.30} & {\scriptsize 11:46:37.99} & {\scriptsize $-$00:11:32.0} & {\scriptsize 2011 Mar} & {\scriptsize 232.734} & {\scriptsize 2.0$\times$1.0} & {\scriptsize 1.2}\tabularnewline
{\scriptsize HATLAS J142413.9$+$022303} & {\scriptsize G15.v2.779} & {\scriptsize 14:24:13.90} & {\scriptsize $+$02:23:04.0} & {\scriptsize 2011 Dec} & {\scriptsize 143.940} & {\scriptsize 4.2$\times$3.1} & {\scriptsize 1.7}\tabularnewline
{\scriptsize HATLAS J090302.9$-$014127} & {\scriptsize SDP.17b$^{*}$} & {\scriptsize 09:03:03.02} & {\scriptsize $-$01:41:26.9} & {\scriptsize 2011 Jan} & {\scriptsize 298.148} & {\scriptsize 2.9$\times$1.1} & {\scriptsize 0.6}\tabularnewline
\hline
\multicolumn{8}{c}{}\tabularnewline
\end{tabular}
\par\end{center}{\scriptsize \par}

\begin{flushleft}
{\small \ \ }\\
{\small a: J2000 coordinates of the centers of the maps displayed in Fig.~2.}\\
{\small b: Central observed frequency (GHz).}\\
{\small c: Total on-source integration time for the PdBI array with six antennas.}\\
{\small {*}: The observations of SDP.17b are reported in Omont et al. (2011).}
\par\end{flushleft}
\end{table*}

\begin{table*}
\caption{Submillimeter and infrared properties of the lensed ultra-luminous galaxies}

\begin{center}
{\small }
\begin{tabular}{lcccccc}
\multicolumn{7}{c}{}\tabularnewline
\hline
\hline
{\footnotesize Source ID} 
& {\footnotesize F$_{250}$} & {\footnotesize F$_{350}$} & {\footnotesize F$_{500}$}
& {\footnotesize $\mu\mathrm{L_{IR}}$} & {\footnotesize $\mu$} 
& {\footnotesize $\mathrm{L_{IR}}$} \tabularnewline
    & {\footnotesize (mJy)} & {\footnotesize (mJy)} & {\footnotesize (mJy)} & {\footnotesize ($\mathrm{10^{13} L}_{\odot}$) } &
&    {\footnotesize ($\mathrm{10^{12} L}_{\odot}$) } \tabularnewline
\hline
{\footnotesize SDP.81} 
& {\footnotesize 129$\pm$20} & {\footnotesize 182$\pm$28} & {\footnotesize 166$\pm$27} & {\footnotesize 5.8} &  {\footnotesize $9.5\pm0.8$} 
& {\footnotesize 6.1}\tabularnewline
{\footnotesize NA.v1.144} 
& {\footnotesize 295$\pm$45} & {\footnotesize 294$\pm$45} & {\footnotesize 191$\pm$31} & {\footnotesize 5.7} & {\footnotesize 5.3$\pm$2.9} 
& {\footnotesize  11} \tabularnewline
{\footnotesize SDP.9} 
& {\footnotesize 485$\pm$73} & {\footnotesize 323$\pm$49} & {\footnotesize 175$\pm$28} & {\footnotesize 4.4} & {\footnotesize $8.5\pm1.8$}  
& {\footnotesize 5.2}\tabularnewline
{\footnotesize NB.v1.78} 
& {\footnotesize 273$\pm$42} & {\footnotesize 282$\pm$43} & {\footnotesize 214$\pm$33} & {\footnotesize 11} & {\footnotesize $10.5\pm1.4$} 
& {\footnotesize 10}\tabularnewline
{\footnotesize G12.v2.30} 
& {\footnotesize 323$\pm$49} & {\footnotesize 378$\pm$57} & {\footnotesize 298$\pm$45} & {\footnotesize 15.7} & {\footnotesize $9.6\pm0.9$} 
& {\footnotesize 16} \tabularnewline
{\footnotesize G15.v2.779} 
& {\footnotesize 115$\pm$19} & {\footnotesize 192$\pm$30} & {\footnotesize 204$\pm$32} & {\footnotesize 8.5} & {\footnotesize $4.1\pm0.2$} 
& {\footnotesize 21}\tabularnewline
{\footnotesize SDP.17b} 
& {\footnotesize 328$\pm$50} & {\footnotesize 308$\pm$47} & {\footnotesize 220$\pm$34} & {\footnotesize 6.9} & {\footnotesize 4.3$\pm$1.2}  
& {\footnotesize 16}\tabularnewline
\hline
\multicolumn{7}{l}{}\tabularnewline
\end{tabular}{\small{} }

\par\end{center}{\small \par}

{\small Note: }{\small 
F$_{250}$, F$_{350}$, 
and F$_{500}$ are the SPIRE flux densities at 250, 350, and 500\,$\rm \mu m$;
$\mu\mathrm{L_{IR}}$ is the apparent total infrared luminosity (8-1000\,$\mu m$); 
$\mu$ is the adopted lensing magnification factor derived from lensing modeling of 880\,$\mu$m continuum, mostly from Bussmann et al.\ in prep. 
(see text Sec.\ 3 for details); 
$\mathrm{L_{IR}}$ is the intrinsic infrared luminosity, derived from the literature (Harris et al.\ 2012; Lupu et al.\ 2012; Fu et al.\ 2012; Bussmann et al.\  2012) except for NB.v1.78, for which an approximate value is inferred from scaling the value of G12.v2.30 with the ratio of the 350\,$\mu$m flux densities. }

{\small \par}

\end{table*}

\begin{table*}
\caption{Observed parameters of the CO emission lines} 
\begin{center}
{\small }%
\begin{tabular}{lcclll}
\multicolumn{6}{c}{}\tabularnewline
\hline 
\hline 
{\footnotesize Source ID} & {\footnotesize $z$}  &  {\footnotesize CO line} & {\footnotesize $\mathrm{~~~I_{CO}}$} & 
{\footnotesize $\mathrm{~~\Delta V_{co}}$} 
& {\footnotesize Other CO data} \tabularnewline
 &  &  & {\footnotesize (Jy.km/s)}
 & {\footnotesize (km s$^{-1}$)} 
 & \tabularnewline
\hline 
{\footnotesize SDP.81} & {\footnotesize 3.040} & {\footnotesize 5-4} & {\footnotesize 7.0$\pm$0.4} & {\footnotesize 560$\pm$40} &
{\footnotesize Harris et al. (2012)} \tabularnewline
{\footnotesize NA.v1.144} & {\footnotesize 2.202} & {\footnotesize 4-3} & {\footnotesize 12.3$\pm$0.6} & {\footnotesize 210$\pm$10} & 
{\footnotesize Harris et al. (2012)} \tabularnewline
{\footnotesize SDP.9} & {\footnotesize 1.574} & {\footnotesize 3-2} & {\footnotesize 9.2$\pm$0.3} &   {\footnotesize 420$\pm$10} & 
{\footnotesize Iono et al. (2012)}     \tabularnewline
{\footnotesize NB.v1.78}  & {\footnotesize 3.111} & {\footnotesize 3-2} & {\footnotesize 3.3$\pm$0.5} & {\footnotesize 560$\pm$70} & 
{\footnotesize Riechers et al. (in prep.)} \tabularnewline
{\footnotesize G12.v2.30} & {\footnotesize 3.259} &  {\footnotesize 3-2} & {\footnotesize 7.8$\pm$1.7} & {\footnotesize 650$\pm$160} & 
{\footnotesize Fu et al. (2012)} \tabularnewline
{\footnotesize G15.v2.779}& {\footnotesize 4.244} &  {\footnotesize 7-6} & {\footnotesize 7.5$\pm$ 0.6} & {\footnotesize 840$\pm$50} & 
{\footnotesize Cox et al. (2011)} \tabularnewline
{\footnotesize SDP.17b} & {\footnotesize 2.305} & {\footnotesize 4-3} & {\footnotesize 9.1$\pm$0.3} & {\footnotesize 320$\pm$10} & 
{\footnotesize Harris et al. (2012)}   \tabularnewline
\hline 
\multicolumn{6}{l}{}\tabularnewline
\end{tabular}{\small{} }
\par\end{center}{\small \par}

\ \ \\
{\small Note: }{\footnotesize $z$}{\small{} is the redshift measured from the PdBI CO spectra (Fig.~1);
$\mathrm{I_{CO}}$ is the integrated flux of the CO line; $\mathrm{\Delta V_{co}}$ is the CO linewidth (FWHM); 
references to other CO measurements from the literature are given in the last column.} {\small \par}
\end{table*}

\begin{table*}
\centering
\caption{Observed parameters of the H$_{2}$O emission lines and continuum emission}
\setlength{\tabcolsep}{3.2pt}
\begin{center}
{\footnotesize}
\begin{tabular}{lccccccccc}
\hline 
\hline 
{\scriptsize Source} & 
{\scriptsize $\mathrm{Line{}^{a}}$} & {\scriptsize $\mathrm{\nu_{obs}}^{b}$} 
& {\scriptsize $\mathrm{S_{\nu} pk\, }^{c}$} & {\scriptsize $\mathrm{S_{\nu}}^{c}$} & {\scriptsize $\mathrm{S_{\nu H_{2}O}pk}^{d}$} 
& {\scriptsize $\mathrm{\Delta V_{H2O}}$} & {\scriptsize $\mathrm{I_{H_{2}O}pk}^{d}$} & {\scriptsize $\mathrm{I_{H_{2}O}}^{d}$} 
& {\scriptsize $\mu\mathrm{L_{H_{2}O}}^e$} \tabularnewline
 & & {\scriptsize [GHz]} & {\scriptsize [mJy/beam]} & {\scriptsize [mJy]} & {\scriptsize [mJy/beam]} & 
{\scriptsize [km s$^{-1}$]} & {\scriptsize [Jy km s$^{-1}$/beam]} & {\scriptsize [Jy km s$^{-1}$]} &
{\scriptsize [$10{}^{7}\,\mathrm{L_{\odot}}$]} \tabularnewline
\hline 
{\scriptsize SDP.81} & 
{\scriptsize 2} & {\scriptsize 244.5000} & {\scriptsize 12.5} & {\scriptsize 27.0$\pm$1.2} & 
{\scriptsize 10.7$\pm 1.9$} & {\scriptsize 140$\pm$50} & {\scriptsize 1.6} & {\scriptsize 1.8$\pm$0.5}  
& {\scriptsize 32$\pm$9} \tabularnewline
{\scriptsize NA.v1.144} & 
{\scriptsize 1} & {\scriptsize 234.8343} & {\scriptsize 9.4} & {\scriptsize 16.4$\pm$0.7} & 
{\scriptsize 22.8$\pm$1.7} & {\scriptsize 200$\pm$50} & {\scriptsize 4.9} & {\scriptsize 7.5$\pm$0.9} & 
{\scriptsize 57$\pm$7} \tabularnewline
{\scriptsize SDP.9} & 
{\scriptsize 1} & {\scriptsize 292.1425} & {\scriptsize 17.4} & {\scriptsize 21.7$\pm$0.9} & 
{\scriptsize 20.3/37.8$\pm$2.9$^f$} & {\scriptsize 410$\pm$50 } & {\scriptsize 11.3} & {\scriptsize 14.4$\pm$1.1} 
& {\scriptsize 60$\pm$5} \tabularnewline
{\scriptsize NB.v1.78} & 
{\scriptsize 2} & {\scriptsize 240.2896} & {\scriptsize 15.4} & {\scriptsize 36.9$\pm$0.4} & 
{\scriptsize 5.0$\pm$0.8} & {\scriptsize 510$\pm$90} & {\scriptsize 2.7} & {\scriptsize 6.7$\pm$1.3}  
& {\scriptsize 122$\pm$24} \tabularnewline
{\scriptsize G12.v2.30} & 
{\scriptsize 2} & {\scriptsize 231.9458} & {\scriptsize 4.9} & {\scriptsize 21.0$\pm$0.8} 
& {\scriptsize 4.5/4.5$\pm$0.7$^f$} & {\scriptsize 690$\pm$80} & {\scriptsize 2.7} & {\scriptsize 6.5$\pm$1.2} & 
{\scriptsize 128$\pm$24} \tabularnewline
{\scriptsize G15.v2.779} & 
{\scriptsize 1} & {\scriptsize 143.3837} & {\scriptsize 5.9} & {\scriptsize 7.3$\pm$0.2} & 
{\scriptsize 3.2/6.6$\pm$0.5$^f$} & {\scriptsize 890$\pm$140 } & {\scriptsize 3.4} & {\scriptsize 4.1$\pm$0.6} 
& {\scriptsize 94$\pm$9} \tabularnewline
{\scriptsize SDP.17b} & 
{\scriptsize 2} & {\scriptsize 298.8646} & {\scriptsize 28.6} & {\scriptsize 37.9$\pm$1.1} & 
{\scriptsize 22.9$\pm$2.6} & {\scriptsize 300$\pm$30} & {\scriptsize 7.2} & {\scriptsize 7.8$\pm$0.9} & 
{\scriptsize 85$\pm$10} \tabularnewline
\hline 
\end{tabular}
\par\end{center}

\begin{flushleft}
{\small a: Lines 1 and 2 correspond to the H$_{2}$O $(2_{11}-2_{02})$ ($\rm \nu_{rest} = 752.03 \, GHz$, $\rm E_{up} = 137 \, K$)
and $(2_{02}-1_{11})$ ($\rm \nu_{rest} = 987.93 \, GHz$, $E_{up} = 101 \, K$) transitions, respectively.}{\small \par}

{\small b: $\rm \nu_{obs}$ is the frequency corresponding to the zero of the velocity scale of the spectra of Fig.\ 1.}
{\small \par}

{\small c: 
$\rm S_{\rm \nu}$pk  corresponds to the peak continuum flux density (mJy/beam); 
$\rm S_{\nu}$ refers to the total spatially integrated continuum flux density (mJy).}{\small \par}

{\small d: $\rm S_{\nu H_2O} \, pk$ corresponds to the peak $\rm H_2O$ flux density (mJy/beam); 
$\rm I_{\rm H_2O} \, pk$ is the peak H$_2$O velocity-integrated flux density (flux in one beam, in Jy $\rm km \, s^{-1}$/beam) and 
$\rm I_{H_2O}$ the spatially integrated H$_2$O line flux (in Jy $\rm km \, s^{-1}$). 
The values derived for SDP.81 
and G12.v2.30. are underestimated (see text).}{\small \par}

{\small e : $\rm \mu L_{H_2O}$ is the apparent luminosity of the observed H$_2$O line using the 
relations  given in, e.g., 
Solomon, Downes \& Radford (1992).}{\small \par}

{\small f : In the cases of SDP.9, G12.v2.30 and G15.v2.779, the $\rm H_2O$ emission lines are 
double-peaked and the parameters in the table refer to double Gaussian fits.}
 
\par\end{flushleft}
\end{table*}

\section{Properties of the $\rm H_2O$ emission lines}

\subsection{General properties}

The seven lensed high-$z$ ULIRGs are all detected with high signal-to-noise ratios ($\rm S/N \ga 6$, except 
 SDP.81, either in the 2$_{02}$--1$_{11}$  or the 2$_{11}$--2$_{02}$ H$_2$O emission line (see Fig.~1), as well as in the underlying 
redshifted submillimeter continuum emission (S/N\,$>$\,20). The water emission lines are strong, with integrated 
fluxes ranging from 1.8 to 14~Jy~km~s$^{-1}$ (Fig.~1 and Table~4). This indicates that in high-$z$ ULIRGs, they are 
the strongest molecular submillimeter emission lines 
after those of high-J CO  (see Sec.\ 4.3; note, however, that 2\,mm and 3\,mm CO lines displayed in Fig.\ 1 may have weaker intensity 
than the 1\,mm $\rm H_2O$ lines). 

The $\rm H_2O$ linewidths (FWHM) range from 140 to $\rm 900 \, km \, s^{-1}$ (Table 4). The lines have a variety of 
profiles including single Gaussian profiles (NA.v1.144, NB.v1.78), double-peaked profiles (G12.v2.779), or asymmetrical profiles
(SDP.9, SDP.17b, SDP.81).  The $\rm H_2O$ and CO line profiles share similar properties (shape, linewidth) in all sources,  
suggesting that there is no strong differential lensing effects between the CO and the $\rm H_2O$ emission lines. 

From the spatially integrated $\rm H_2O$ line flux, $\rm I_{\rm H2O}$, the apparent $\rm H_2O$ luminosity, 
$\rm \mu L_{H_2O}$ where $\mu$ is the lensing magnification factor, can be derived using the 
relations given in, e.g., 
Solomon, Downes \& Radford (1992). 
$\rm \mu L_{H_2O}$ varies by only a factor of $\sim 4$ from SDP.81 to NB.v1.78 or G12.v2.30 (Table~4).  Fig.~3 displays 
the relation between $\rm \mu L_{H_2O}$ and  $\rm \mu L_{IR}$, including the values for  local ULIRGs (Yang et al.\ in prep.). 
The infrared and H$_2$O apparent luminosities are one or two orders of magnitude higher in these high-$z$ galaxies than in local ULIRGs. However, the ratios $\rm L_{H_2O}/L_{IR}$ are similar, although slightly higher for the high-$z$ galaxies.

Figure~2 displays the images of the $\rm H_2O$ emission line and the corresponding submillimeter continuum emission for the
seven lensed high-$z$ ULIRGs. The relatively low angular resolution (at best $\sim 1^{\prime\prime}$) of the current data  limits 
for most of the sources a detailed study of the spatial properties of their signal. Four of the sources remain unresolved 
at the present angular resolution: SDP.9, SDP.17b, NA.v1.144, and G15.v2.779. 
The three remaining sources are extended and display distinct lensed morphologies including two-image
configuration system
(SDP.81), a complex elongated structure (G12.v2.30), and an  extended structure (NB.v1.78).  
Each source is described in detail below. 

Table 4 reports the values of the H$_2$O velocity-integrated flux density (line flux) detected  at the peak of the source in one synthesized beam, 
$\mathrm{I_{H_{2}O}pk}$, and the velocity-integrated line flux integrated over the source size, $\rm I_{H_2O}$. In most cases, the ratio 
$\rm I_{H_2O}/I_{H_2O}$pk  is lower than 1.5, except for the extended sources G12.v2.30 and NB.v1.78. 
For the other sources studied in this paper, except for SDP.81, it is thus unlikely that significant flux is missed. 
This is also confirmed 
 when comparing our total spatially integrated continuum flux density $\rm S_{\nu}$ with measurements 
published elsewhere at the nearby frequency of 250~GHz, using the bolometer MAMBO camera at the
IRAM 30-meter telescope (Negrello et al.\ 2010; Dannerbauer et al.\ in prep.). NA.v1.144  and SDP.81, 
which were observed at the PdBI close to 250\,GHz, agree very well with MAMBO observations. For SDP.9 and SDP.17b, which were observed 
at the PdBI close to 300\,GHz, the comparison is more difficult; however, $\rm S_{\nu}$ is somewhat higher than the values 
expected from extrapolations of the MAMBO values. Although there is no MAMBO observation for NB.v1.78, $\rm S_{\nu}$ 
agrees fairly well with the flux density measured at 880\,$\mu$m using the Smithsonian Millimeter
Array (SMA) (Bussmann et al.\ in prep.). Finally, the continuum flux density measured at 143\,GHz for G15.v2.779 is in 
excellent agreement with the value reported by Cox et al.\ (2011) at 154\,GHz.
 
The 3.6\,GHz bandwidth covered by the PdBI receivers (and correlator), which was generally centered on the  
redshifted frequency of the H$_2$O emission line, encompasses other potential emission lines of interest,  
including i) $^{13}$CO, H$_2^{18}$O, H$_2^{17}$O, H$_3$O$^+$ in the vicinity of 
H$_2$O(2$_{02}$--1$_{11}$); and ii) H$_2$O$^{+}$(2$_{02}$--1$_{11}$,J$_{3/2,3/2}$ and J$_{5/2,3/2}$)  in the 
vicinity of H$_2$O(2$_{11}$--2$_{02}$)). However, the current sensitivity of our spectra 
is too low to detect these lines, which are not seen in the {\it Herschel} spectra of Mrk\,231 and Arp\,220 
except for the two H$_2$O$^+$ lines, which are present in the spectrum of  Arp\,220 (Rangwala et al.\ 2011). 
Apart from the H$_2$O emission lines, no other emission line is detected at the S/N ratio of the 
spectra of the seven high-$z$ galaxies studied in this paper.

\subsection{Individual sources} 

The following describes the properties of the individual lensed galaxies. 
 
\subsubsection{SDP.81 at $z\,=\,3.040$}

The source SDP.81 is at the highest redshift among the SDP lensed galaxies of Negrello et al.\ (2010). 
It displays two arcs in the SMA 880~$\rm \mu m$ data, which are also seen in the {\it HST} (Negrello et al. in prep.) and the PdBI CO imaging results (Cox, Ivison et al. in prep.).  
Preliminary lensing modeling indicated a very high lensing magnification 
factor of $\mu$\,=\,19 (Negrello et al.\ 2010), but this value has been revised downward to 
$\mu$\,=\,9.5$\pm$0.5 based on recent {\it HST} data (Dye et al. in prep.). This is confirmed by the SMA 880~$\rm \mu m$  imaging results which yield $\rm \mu = 9.5\pm0.8$ (Bussmann et al. in prep. -- hereafter B13). 
As discussed in Sec.\ 4.1, we adopted the values of $\rm \mu$ derived from the 880~$\rm \mu m$ SMA data by B13 as basic reference for the lensing magnification factors listed in Table~2.

The CO line profile is asymmetrical with a red component that is much stronger than the blue one.  
The line profile of the $\rm H_2O$ emission line is also dominated by a main 
red component (peaking at $\rm \sim +200 \, km \, s^{-1}$ with a width
of  $\rm \Delta v = 140 \, km \, s^{-1}$). Due to the low S/N ratio of the $\rm H_2O$ spectrum, the 
blue emission component of the profile (peaking at $\rm -100 \, km \, s^{-1}$) that is clearly seen 
in the CO spectrum is only barely visible in the $\rm H_2O$ emission spectrum (Fig.~1). 

Despite the low angular resolution, the dust continuum image is clearly resolved  (Fig.\ 2), showing both 
gravitational arcs of the 880\,$\mu$m image of Negrello et al. (2010), with a stronger emission in the eastern arc. This structure is also seen in the CO map of  Cox, Ivison et al.\ which indicates that  
the stronger red emission traces the bright eastern arc,  whereas the weaker blue emission 
corresponds to the western arc. 
Due to the low S/N ratio of the data, the H$_2$O emission is seen only in the eastern arc, 
consistent with the only clear detection of the red part of the spectrum. 

The $\rm H_2O$ 
line flux I$_{\rm H2O}$ and the apparent L$_{\rm H_2O}$ luminosity, which are given in Table~4, take into account  
the total emission including the red and blue components. However, due to the low S/N ratio of 
the $\rm H_2O$ data, these numbers are less accurate than for most other sources in our sample. 
Because the emission is very extended, the spatially integrated H$_2$O line flux $\rm I_{H_2O}$ is probably 
underestimated, as indicated by the absence of significant H$_2$O emission seen in the western arc, and the very small difference between $\rm I_{H_2O}$ and $\rm I_{\rm H_2O}pk$ compared to the continuum ratio $\rm S_{\nu}$/$\rm S_{\rm \nu}$pk\,=\,2.2.

\subsubsection{NA.v1.144 at $z\,=\,2.202$}

The narrow $\rm H_2O$ linewidth of NA.v1.144 is the same as is seen in the CO emission line 
($\rm \Delta v \sim 200 \, km \, s^{-1}$). It is the narrowest CO linewidth of our sample (Table~3).  

The estimate of the lensing magnification factor from detailed modeling based on the SMA image of the 880~$\rm \mu m$ continuum emission indicates a low value of $\mu \sim 4.6 \pm 1.5$ (B13). 
However, the SMA imaging of this object has limited uv coverage (only the extended array was used) 
which might compromise the measurement of the lensing magnification factor. The corresponding values for the IR and 
CO luminosities appear to be surprisingly high compared to the narrow linewidth. Indeed, the uncertain derivation 
of $\mu$ from the CO linewidth and 
luminosity (Harris et al.\ 2012) yields $\mu_{\rm CO} \sim 23\pm15$. 
Therefore, we have 
doubled the uncertainty on the SMA value of $\mu$ and  adopted  the rms$^2$-weighted 
average of the estimate from the SMA dust-continuum and the CO line, i.e. $\mu$ = 5.3$\pm$2.9. 

This bright submillimeter galaxy is clearly unresolved with our present angular 
resolution, a result that is confirmed by the small extension observed in the SMA results (B13).

\subsubsection{SDP.9 at $z\,=\,1.574$}

The galaxy SDP.9 is the brightest $250 \, \rm \mu m$ and lowest redshift of the five lenses discussed in 
Negrello et al. (2010). Its lensing magnification factor is estimated to be $\mu$\,=\,8.7$\pm$0.7 from {\it HST} imaging 
(Dye et al.\ in prep.).  We adopted here the SMA 880\,$\mu$m value from B13, $\mu$\,=\,8.5$\pm$1.8, which agrees with the HST value and also with 
$\rm \mu_{CO} = 11$. 

The $\rm H_2O$ intensity of SDP.9 is by far the largest of our sample, due to the low redshift of the source.
The corresponding  apparent $\rm H_2 O$ luminosity is among the highest found so far. The emission line 
profiles in H$_2$O and CO are similar, including the strong red emission spike and the weaker, broader blue 
emission (Fig.~1). The high angular CO data from Cox, Ivison et al.\ (in prep) 
as well as the {\it HST/WFC3} imaging data (Negrello et al. in prep.)  
show that SDP.9 is a good example of 
an Einstein ring, with a clear asymmetry between the east and west arcs.  However, our angular resolution is 
not high enough to 
resolve the emission. 

\subsubsection{NB.v1.78 at $z\,=\,3.111$}

This source has one of the highest redshifts and highestt apparent luminosities in our sample. 
The intermediate
linewidth of the CO and H$_2$O lines ($\rm \Delta v \sim 550 \, km \, s^{-1}$) is consistent with those measured
in the mid-J CO lines used to initially determine the redshift with
CARMA (Riechers et al.\ in prep.). The H$_2$O line
is indeed the brightest line measured in this source so far and thus
provides the best constraints on the line profile.
The moderate CO intensity and linewidth indicates intermediate CO, H$_2$O and infrared intrinsic luminosities, 
L$_{\rm IR}$\,$\approx$\,$10^{13} \, \rm L_{\sun}$. 

The lensing magnification factor, derived from the SMA 880\,$\mu$m imaging results, is $\mu$ = 10.5$\pm$1.4 (B13), a 
value which is adopted in this paper and is consistent with $\mu_{\rm CO} = 12$ (inferred from the CO data of Table 3).

The source is resolved and shows a relatively complex morphology in the $\rm H_2O$ and continuum maps.
The peak value in a single beam, $\mathrm{I_{H_{2}O}}$pk, reported in Table~4, is  lower than the spatially integrated value 
I$_{\rm H_2O}$ by a factor $\sim$2.5. The ratio is similar for the 240~GHz continuum, where the peak value is 
15.4~mJy/beam and the integrated flux density is 36.9~mJy (see Table~4).  
In addition to the southeast main peak, there are northern and southwestern extensions (Fig.~2). 
This morphology is also seen in recent SMA $\rm 880 \, \mu m$ imaging results (Bussmann et al.\ in prep.). 
However, the southwest peak is very strong in H$_2$O compared to the main southeast peak and the peak ratio 
in the continuum. This could  suggest that the H$_2$O emitting region is more compact than the continuum. 

\subsubsection{G12.v2.30 at $z\,=\,3.259$}

G12.v2.30 is another prominent source among the H-ATLAS brightest lenses.  
Its complex differential lensing has been analyzed by Fu et al. (2012), showing that the lensing is very 
different in the near-infrared compared to the submillimeter or CO emission.   The various estimates of the lensing magnification 
factor based on submillimeter images or the CO linewidths are comparable:  
$\mu$\,=\,7.6$\pm$1.5 for the dust submillimeter and 6.9$\pm$1.6 for $\rm CO(1-0)$ imaging by Fu et al.; 
9.6$\pm$0.9  for the same submillimeter data by B13 (the value adopted here); 
$\mu_{\rm CO}$\,=\,7$\pm$2 by Harris et al.\ (2012) from $\rm CO(1-0)$ 
linewidth and luminosity. 
These values confirm that G12.v2.30 is one of the most luminous infrared galaxies in H-ATLAS with 
an estimated $\rm L_{IR} \sim 1.6 \times 10^{13} \, L_{\sun}$.

The H$_2$O profile appears to be double peaked with a total width of $\sim$\,700 km/s. Due to the short integration time of the observations, the CO results for G12.v2.30 are the noisiest of our sample and the details of the profile, in particular, its double-peaked nature remain unclear. Within the noise, the CO profile is compatible with the H$_2$O emission, although additional observations are needed to confirm this result.

The source is resolved with the present angular resolution of $2.0'' \times 1.0''$ and shows a complex structure that 
is most clearly seen in the continuum map (Fig.~2): an elongated arc to the south shows two peaks and emission is seen 
about $2''$ to the north. Due to the low S/N ratio, only the southern component is detected in the $\rm H_2O$ 
emission line.  G12.v2.30 appears to be a complex source. The dust and gas emissions as well as the 
stellar emission have very different morphologies, which indicates that the lensing potential is unusual (Fu et al.\ 2012).
Estimating the total intensity of the H$_2$O line remains difficult for G12.v2.30.  
The peak value in a single beam, $\mathrm{I_{H_{2}O}}$pk, reported in Table~4, is lower than the spatially integrated value 
I$_{\rm H_2O}$ by a factor $\sim$2.4. The situation is similar for the 240~GHz continuum, where the peak value is 
4.9~mJy/beam and the integrated flux density is 21.0~mJy (see Table~4). This latter value is itself much lower than the flux density 
of $\rm \sim 36 \, mJy$ measured at 1.2\,mm using MAMBO at the 30-meter telescope (Dannerbauer et al.\ in prep.). 
However, the reported value for the spatially integrated line flux I$_{\rm H_2O}$ is less accurate than for most other sources in our sample. 
Because the emission is very extended, the spatially integrated H$_2$O line flux $\rm I_{H_2O}$ is probably 
underestimated, similarly to the continuum, as indicated for the latter by the factor $\sim$1.5 lower compared to the MAMBO value.  

\subsubsection{G15.v2.779 at $z\,=\,4.244$}
With a redshift higher than 4 and F$_{500\mu m}$\,=\,204$\pm$32\,mJy, G15.v2.779 (also refered to as ID.141) remains unique 
in the H-ATLAS survey (Cox et al.\ 2011). Its lensing magnification factor has recently been estimated to be only $4.1\pm0.2$  
(Bussmann et al.\ 2012), which makes it the most luminous H-ATLAS source, with 
$\rm L_{IR} \, = \, 2 \times 10^{13} \, L_{\sun}$. This is consistent with the broad 
CO lines, the widest of our sample ($\rm 760-890 \, km \, s^{-1}$), which would imply $\rm \mu_{CO} \sim 8$. 
The $\rm H_2O$ line  has one of the 
largest S/N  thanks to good observing conditions in the 2\,mm window. 

The width and the characteristics of the complex, double-peaked $\rm H_2O$ line agree well  
with those of the CO(7-6) and [C{\small I}] lines (Fig.\ 1). 
G15.v2.779 has the highest intrinsic $\rm H_2O$ luminosity of our sample by a factor $\sim$2. This implies that both should have very large $\rm H_2O$ luminosities if the two components corresponded to two galaxies. 

At the present angular resolution G15.v2.779 remains unresolved, in agreement with the compactness of the CO 
(Cox, Ivison et al.\ in prep.) and the 880\,$\mu$m  continuum emissions (Bussmann et al.\ 2012).

\subsubsection{SDP.17b at $z\,=\,2.305$}

The  detection of the $\rm H_2O$ emission line in SDP.17b was reported and discussed in Omont et al.\ (2011). 
Its $\rm H_2O$ line profile is in perfect agreement with the CO emission profile - see Fig.~1 for the CO(4--3); 
also Harris et al.\ (2011) for the CO(1--0) and Iono et al.\ (2012) and Riechers et al.\ in prep.\ for the CO(3--2) profiles - including 
the asymmetry with the distinct blueshifted  emission wing. The source is not resolved in the present data and 
higher angular resolution data indicate that the source remains compact and displays a velocity gradient (Cox, Ivison
et al.\ in prep.). 

Again the $\mu$ value derived by B13, $\mu$ = 4.0$\pm$0.6, is much lower than approximately expected from the CO linewidth and 
luminosity, which yields  $\mu_{\rm CO}$ = 18$\pm8$ (Harris et al.\ 2012) - note that this latter value is uncertain because of the asymmetric  CO line profile. 
However,  the image separations are at 
the limit of what can be resolved by the SMA. Therefore, we have again 
 doubled the uncertainty on the SMA value of $\mu$ and  adopted  the rms$^2$-weighted  
average of the estimate from the SMA dust-continuum and the CO line, i.e. $\mu$ = 4.3$\pm$1.2. 
However, the line of sight toward SDP.17b could include an intervening system, as indicated by a CO line 
 at $z=0.942$ reported by Lupu et al.\ (2012) based on Z-Spec data, in addition to the $z=2.3$ line seen by the GBT 
(Harris et al.\ 2012) and the PdBI (Fig.\ 1). But this  CO line at $z=0.942$ was not confirmed by Omont et al.\ (2011)
and more observations are needed to clarify this question.  If there was such a dusty object at $z=0.942$, 
the lens model derived from the SMA data
 of this very compact source would have to be revised.

\begin{table*}
\centering
\caption{Properties of the dust continuum and the H$_{2}$O line emission}

\begin{center}
{\small }%
\begin{tabular}{lcccccc}
\multicolumn{7}{c}{}\tabularnewline
\hline
\hline
{\small Source ID} 
& {\small line$^{a}{}_{obs}$}  & {\small $\mathrm{L_{IR}}$} & {\small $\mathrm{L_{H_{2}O-1m}}$} & {\small $\mathrm{L_{H_{2}O-1a}}$} & {\small $\mathrm{L_{H_{2}O-2m}}$} & {\small $\mathrm{L_{H_{2}O-2a}}$}\tabularnewline
  &  &  {\small ($10{}^{12}\,\mathrm{L_{\odot}}$)} & {\small ($10{}^{7}\,\mathrm{L_{\odot}}$)} & {\small ($10{}^{7}\,\mathrm{L_{\odot}}$)} & {\small ($10{}^{7}\,\mathrm{L_{\odot}}$)} & {\small ($10{}^{7}\,\mathrm{L_{\odot}}$)}\tabularnewline
\hline
{\small SDP.81}    
& {\small 2} & {\small 6.1}  & {\small(1.5)$^b$} & {\small(2.3)}  & {\small 3.3}   & {\small  3.3}\tabularnewline
{\small NA.v1.144} 
& {\small 1} & {\small 11}  & {\small 9.7}  & {\small 9.7}   & {\small(25)} & {\small(16)}\tabularnewline
{\small SDP.9}     
& {\small 1} & {\small 5.2}  & {\small 7.0}  & {\small 7.0}   & {\small(16)} & {\small(10)}\tabularnewline
{\small NB.v1.78}  
& {\small 2} & {\small 10} & {\small(5.1)} & {\small(7.9)}  & {\small 12}  & {\small 12}\tabularnewline
{\small G12.v2.30} 
& {\small 2} & {\small 16} & {\small(5.8)} & {\small(9.0)}  & {\small 13}  & {\small 13}\tabularnewline
{\small G15.v2.779}
& {\small 1} & {\small 21} & {\small 23} & {\small 23}  & {\small(55)} & {\small(34)}\tabularnewline
{\small SDP.17b}   
& {\small 2} & {\small 16} & {\small(8.7)} & {\small(14)} & {\small 20}  & {\small 20}\tabularnewline
\hline
\multicolumn{7}{l}{}\tabularnewline
\end{tabular}
\par\end{center}{\small \par}

\begin{flushleft}
{\small a: Line 1 is H$_{2}$O$(2_{11}-2_{02})$ 752.03GHz and line
2 is H$_{2}$O$(2_{02}-1_{11})$ 987.93GHz;}\\
{\small b:  Luminosity values for unobserved lines are written in parenthesis, they are inferred from the other observed lines assuming
 the line intensity ratio from Mrk 231 (Gonz$\acute{a}$lez-Alfonso et
al. 2010, subscript m) or Arp220 (Rangwala et al. 2011, subscript
a). As detailed in the text, we consider that these values significantly underestimate the H$_2$O luminosity for SDP.81 and G12.v2.30.
}
\par\end{flushleft}
\end{table*}{\small{} }

\section{Discussion}

\subsection{Lensing}

\pagestyle{empty}
\setcounter{secnumdepth}{3}
\setcounter{tocdepth}{3}
\makeatletter
\makeatother
\begin{figure}
\begin{center}
{\small \includegraphics[scale=0.85]{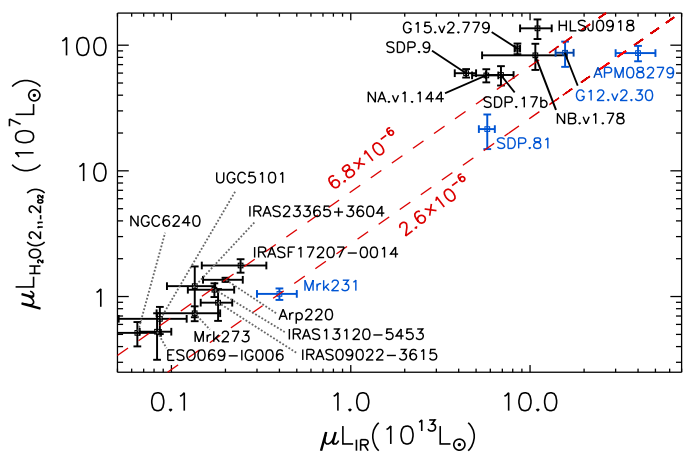}} 
\caption{Apparent H$_{2}$O luminosity versus apparent infrared luminosity of high-redshift sources, for H$_2$O 
Line 1 ($2_{11}-2_{02}$, 752.03\,GHz) with the Arp\,220 H$_2$O line ratio (see text), compared to local ULIRGs ($\mu$L$_{\rm IR}$\,$<$\,4 $\times 10^{12} \, L_{\sun}$, Yang et al.\ in prep.). The 
dashed diagonal lines have constant L$_{\rm H2O}$/L$_{\rm IR}$ ratios 6.8\,x\,10$^{-6}$ and 
2.6\,x\,10$^{-6}$, equal to those of Arp\,220 and Mrk\,231.  Because of their peculiarities, values  for SDP.81 and G12.v2.30 (underestimated by factors possibly up to $\sim$1.5-2 because they are very extended, see text), 
and those of APM\,08279+5255 and Mrk\,231 (type 1 and type 2 QSOs, respectively)  are printed in blue.  
} 
\par\end{center}
\end{figure}

The main purpose of this paper is to investigate the importance of  the H$_2$O emission in high-$z$ ULIRGs, but not to study 
its spatial structure by separating it from lensing  images. This is the reason why the observations were carried out 
with the most compact configuration of PdBI to maximize 
the sensitivity for detection. 
Detailed lensing modeling is thus out of the scope of this paper. However, a good control of lensing effects is mandatory 
for various reasons: i) 
As seen above, the H$_2$O emission could be resolved out in sources that are extended, even with the limited resolution of the observations reported in this paper. 
ii) Differential lensing 
effects must be controlled for comparison with emission from other H$_2$O lines or other molecules or dust.  
 As discussed, e.g., by Serjeant (2012), the importance of differential lensing is enhanced for the 
highest magnifications which take place close to caustics, and, especially, for blank fields, such as H-ATLAS where the 
deflectors are either single massive galaxies or small groups of galaxies.
A striking example is the large difference observed in lensed images in the near-infrared compared to the submillimeter or CO emission of G12.v2.30 (Fu et al.\ 2012). 
iii) Lensing magnification must be corrected for deriving intrinsic properties of the sources such as luminosities 
or line ratios.

To infer intrinsic infrared and H$_2$O luminosities of our sources, a first approximation is  to derive a (mean) lensing magnification factor $\mu$. We  systematically preferred values of $\mu$ inferred from submillimeter imaging rather than near-infrared {\it HST} imaging, since H$_2$O emission 
should be more related to submillimeter dust emission than to near-infrared stellar emission, and submillimeter and near-infrared lensing can 
be very different as examplified in the case of G12.v2.30 (Fu et al.\ 2012).
As discussed in Sec.\ 3.2, the lensing magnification factor $\mu$ is thus determined from existing 
lensing models based on 880\,$\mu$m SMA  imaging results (B13; see also Bussmann et al.\ 2012). For two sources these values were slightly corrected by also using the values of $\mu$ derived following a method described in Harris et al. (2012), which uses the relation bewteen the CO
linewidth and luminosity.  
 As the low angular resolution and limited S/N ratio of our H$_2$O data do not allow us to trace possible differential effects of lensing, we adopted  a minimum uncertainty of $\pm$20\% for all values of $\mu$.

\subsection{Relation between H$_2$O and infrared luminosities}

\pagestyle{empty}
\setcounter{secnumdepth}{3}
\setcounter{tocdepth}{3}
\makeatletter
\makeatother
\begin{figure*}
\begin{center}
{\small \includegraphics[scale=1.0]{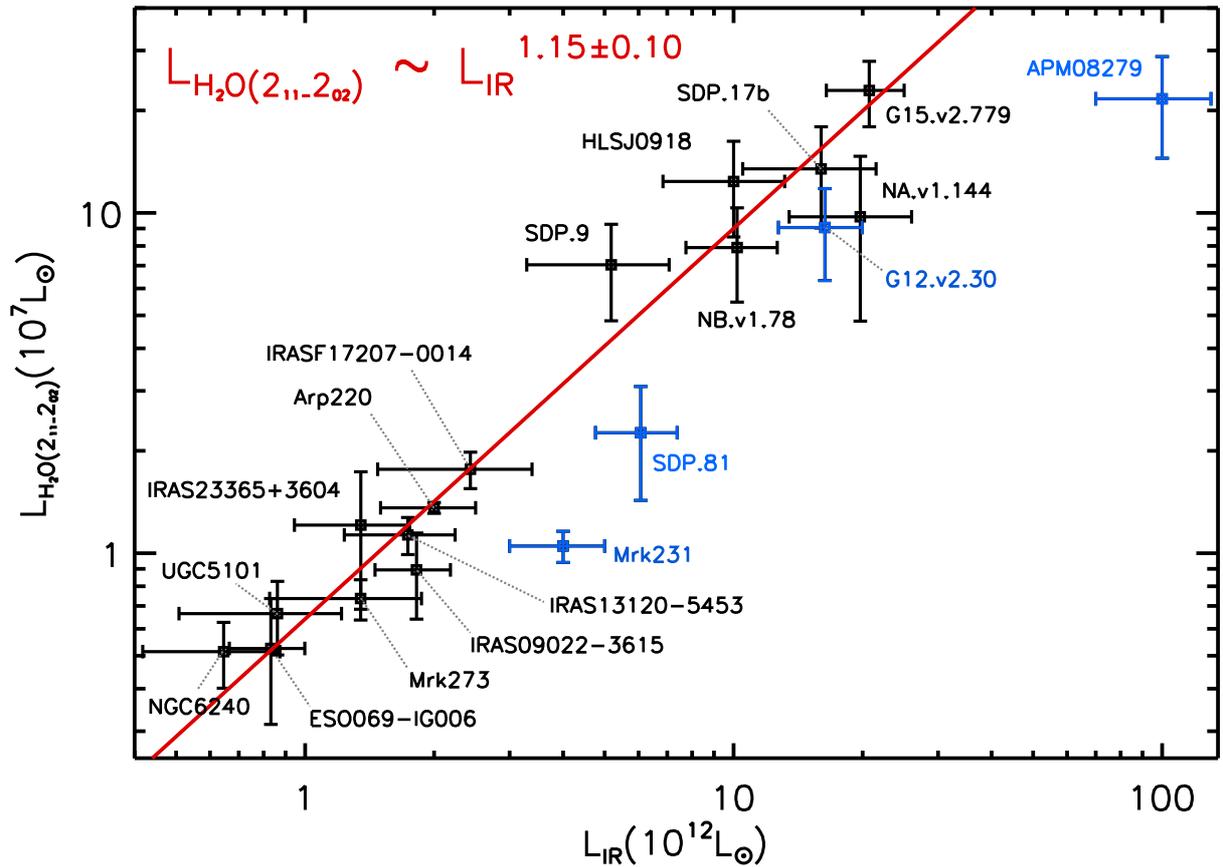}}
\caption{Relation of the intrinsic H$_{2}$O luminosity with the infrared luminosity for the seven ultra-luminous infrared galaxies 
reported in this paper, other well-known high-$z$ sources taken from the literature, and local ULIRGs (L$_{\rm IR}$\,$<$\,4 $\times 10^{12} \, L_{\sun}$, Yang et al.\ in prep.). The luminosity is reported to the 
H$_2$O ($2_{11}-2_{02}$, 752.03\,GHz) emission line (referred to as 'line 1' in this paper), where we adopted an 
Arp\,220 H$_2$O line ratio when another $\rm H_2O$ line ('line 2') was measured (see text). The solid line shows a best-fit power-law, 
$\rm L_{H_2O} = L_{IR}^\alpha$, where $\alpha = 1.17 \pm 0.08$. 
Error bars are defined in the following way: 
i) for L$_{\rm IR}$, they combine 
the uncertainties on the values of $\mu$\,L$_{\rm IR}$ and the uncertainties on $\mu$ given in Table 2; 
ii) for $\rm L_{H_2O}$, they include the errors on $\rm I_{H_2O}$ of Table 4, and the same uncertainties on $\mu$ except that when the latter is <20\% it is replaced by 20\% (see Sec.\ 4.1).
Because of their peculiarities, values (in blue) for SDP.81 and G12.v2.30 (both underestimated by factors possibly up to $\sim$1.5-2 because they are very extended, see text), and those of APM\,08279+5255 and Mrk\,231 (type 1 and type 2 QSOs, respectively) are not taken into account for the fit. 
}
\par\end{center}
\end{figure*}

The intrinsic H$_2$O luminosities, based on the magnification factors listed in Table 2, are reported for each source in Table 5. The information on the H$_2$O emission is limited by the observation of a single emission 
line per source. This prevents us from comparing the H$_2$O excitations  between different sources. 
It is nevertheless interesting to compare their H$_2$O line luminosity and to check their dependency on other properties, 
such as the infrared intrinsic luminosity, $\rm L_{IR}$, since infrared excitation is thought to play a key role in the 
excitation of the H$_2$O levels. 

In addition, the $\rm H_2O$ lines that were observed differ from 
source to source: three sources were measured in the H$_2$O(2$_{11}$--2$_{02}$) emission line 
('Line 1')
and four sources in the H$_2$O(2$_{02}$--1$_{11}$) emission line ('Line 2')
To deal with this disparity, we 
approximated the integrated line flux, I$_{\rm H_2O}$,  of the emission line that was not observed,  by adopting
the value of the line ratio of 'Line 2' to 'Line 1' measured either in Arp\,220, r$_{21}$\,=\,3440/2970\,=\,1.16 (Rangwala et al.\ 2011) or in Mrk\,231, r$_{21}$\,=\,718/415\,=\,1.73 
(Gonz\'alez-Alfonso et al.\ 2010), 
listed as 'a' and 'm' in Table~5. 
The corresponding values of  L$_{\rm H_2O}$ are reported in parenthesis in Table~5. 
Finally, 
the full set of values of the intrinsic 
luminosities L$_{\rm H_2O-1/2}$ for lines 1 and 2 and the 
line ratios of Arp\,220 or Mrk\,231
are reported in Table~5. 

There are several arguments to favor the conversion factor of Arp\,220 rather than Mrk\,231: for all galaxies, except G12.v2.30, the CO excitation seems significantly lower than in Mrk\,231 (Lupu et al.\ 2012
); using the Mrk\,231 conversion factor seems to yield anomalously low values for the line 1 luminosity of sources only observed in line 2 (Table~5).

Therefore, keeping all uncertainties in mind, the relationship 
between $\rm L_{H_2O}$ and $\rm L_{IR}$ is displayed in Fig.\ 4 for $\rm L_{H_2O-1a}$ (line 1 with Arp\,220 line ratio). 
This relationship is inferred from all sources studied in this paper - with the exception of SDP.81 and G12.v2.30 
whose H$_2$O intensities are uncertain and probably underestimated by factors possibly up to $\sim$1.5-2 because they are very extended - together with HLSJ0918 from Combes et al.\ (2012), 
and local ULIRGs (L$_{\rm IR}$\,$<$\,4 $\times 10^{12} \, L_{\sun}$) from  Yang et al.\ (in prep.). 
A best-fit power-law to this relation following

\begin{equation}
 L_{\rm H_2O} = L_{\rm IR}^\alpha
\label{eq1000}
\end{equation}

yields $\alpha$\,=\,1.17$\pm$0.0.10 using 'line 1' and the Arp~220 line ratio. Using the Mrk\,231 line ratio would yield $\alpha$\,=\,1.11$\pm$0.10, whereas
using 'line 2' would give $\alpha$\,=\,1.22$\pm$0.10  and 1.28$\pm$0.12 for the Arp\,220 and
Mrk\,231 line ratios, respectively. Independent of the selected lines or ratios, these results indicate that $\rm L_{H_2O}$ 
increases rapidly with $\rm L_{IR}$, possibly faster than linear. We note, however, that since systematic errors were not taken into account in this analysis, adding the statistical and systematic errors could result in a relation that is closer to being linear. A steep increase is consistent with a significant role of infrared radiation in the excitation of the water vapor lines.
It is interesting to note 
that a similar behavior with the infrared luminosity has been found for the intensity of local H$_2$O and OH megamasers, 
which could be interpreted as a result of the importance of infrared pumping on these masers (e.g.\ Lo 2005).

However, it is important to stress that, as shown e.g. by the modeling of H$_2$O excitation in Mrk\,231 by Gonz\'alez-Alfonso et al., the intensity of (optically thick) H$_2$O lines results from a complex interplay between various parameters that includes gas temperature and density, H$_2$O abundance, infrared radiation field, spatial distribution of H$_2$O and dust, etc. 
The present data, with only one H$_2$O emission line per source, prevents us for constraining the water excitation. Multi-line H$_2$O observations are required for a detailed modeling of the excitation. Therefore, we defer the study of the water excitation in these high-z sources to a future paper, which will report measurements of multiple water transitions. 

The high--J HCN lines seem to display a similar behavior with the infrared luminosity, with $\rm L_{HCN}$ growing possibly 
faster than linear with $\rm L_{IR}$. For instance, Bussmann et al.\ (2008) reported  that $\rm L_{HCN(3-2)}$ varies 
as $\rm L_{IR}^{1.3\pm0.1}$ in a sample of 30 nearby galaxies, although correcting for the difference in emission 
extension seems compatible with a linear relation (Gao et al.\ in prep.). The strong $\rm HCN(5-4)$ emission observed by Weiss et al. (2005) in APM\,08279+5255  seems to require infrared excitation. 
The high HCN 6-5/5-4 ratio observed by Riechers et al.\ (2010) ultimately confirmed that infrared excitation is important for the HCN excitation in this source.
Note that this is not true for 
$\rm HCN(1-0)$, which gives a strictly linear relation, with an index of 1.0, consistent with the interpretation  
that $\rm HCN(1-0)$ is proportional to the mass of dense gas (Solomon et al.\ 1992; Gao \& Solomon 2004). 
However, because the H$_2$O and high-J HCN line intensities may
be increasing faster than linear with $\rm L_{IR}$ 
may indicate the importance of IR excitation. 
This needs to be confirmed by better statistics and by future observations
of higher transitions.

\subsection{Comparison of H$_2$O lines with high-J CO lines and the continuum}

In all cases where the information is available on adjacent CO lines - CO(6-5) or (7-6) for H$_2$O(2$_{11}$--2$_{02}$) (line 1), 
and CO(8-7) or (9-8) for  H$_2$O(2$_{02}$--1$_{11}$) (line 2) - the line flux I$_{\rm H2O-1/2}$ is at least $\sim$0.3--0.5 that of the 
neighboring high--J CO lines. The lack of detailed data on high--J CO emission for most of the 
sources limits any detailed comparative study.

The three SDP sources and G12.v2.30 benefit from a complete spectral coverage from 200 to 300\,GHz with 
ZSpec (Lupu et al.\ 2012 and in prep.). This range includes 2 to 5 high--J CO lines, depending on the redshift. 
Even though the sensitivity and the spectral resolution of ZSpec are limited, these data allow us to derive useful 
trends on the strength of CO as the approximate spectral line energy distributions (SLEDs) derived by Lupu et al.\ (2012). 

In SDP.17b, the ratio I($\rm H_2O (2_{02}-1_{11})$)/I($\rm CO(8-7)$) is about 0.5 (Omont et al. 2011). 
For SDP.9, information  is lacking on CO emission from levels $\rm J > 6$ because of the low redshift ($z=1.57$)
of the source. However, the CO(6--5) emission line is strong 
with a ratio  I($\rm H_2O (2_{11}-2_{02})$)/I($\rm CO(6-5)$)$\sim$0.36. 
In the case of SDP.81, the H$_2$O and CO emission lines are both weak. Accounting for the part of the H$_2$O 
emission that is not detected in the intrinsically weaker western arc (see Sect. 3.2.1),
we estimate that the ratio  I($\rm H_2O (2_{02}-1_{11})$)/I($\rm CO(7-6)$) is $\approx$ 0.3 - 
note that the CO J\,$>$\,7 lines are not detected in the ZSpec data reported by Lupu et al. (2012).  
The ZSpec spectrum of G12.v2.30 displays strong lines of H$_2$O(3$_{21}-3_{12}$) and CO(10--9), with about
similar intensities  (Lupu et al.\ in prep.), but their ratio may be uncertain by a factor up to two 
because of the noise  in this frequency range of ZSpec.

The best-documented comparison of H$_2$O and CO lines is found for G15.v2.779 because of its 
high redshift ($z=4.24$) and the sensitive PdBI CO observations of Cox et al. (2011). For G15.v2.779, the 
ratio I($\rm H_2O (2_{11}-2_{02})$)/I($\rm CO(7-6)$\,=\,0.52. For the two other sources, NA.v1.144 and NB.v1.78, 
the lack of information on the high-J CO levels prevents any comparison with the H$_2$O emission lines. 

Moreover, the 
intensities of H$_2$O lines ($\rm I_{\rm H2O}$ - Table 4) and of the lower-J CO 2\,mm and 3\,mm lines (Table 3 and Fig.\ 1) 
are roughly comparable. 
Another direct comparison is the relation between the intensity of the H$_2$O emission line (hence the H$_2$O luminosity) 
with the underlying continuum emission (hence the infrared luminosity) - see Table 4 and Figs.\ 3 and 4. But the comparison 
between I$_{\rm H_2O}$ and the continuum flux density S$_{\rm \nu}$ significantly depends on the redshift and the rest 
frequency. 
Finally, as shown by Fig.\ 1 and Table 4, the H$_2$O and continuum flux densities, 
e.g.\ S$_{\rm \nu H_2O}$pk and S$_{\rm \nu}$pk, are roughly comparable, although their ratio 
obviously depends on  the linewidth. 

\subsection{Implication of ubiquitous detection of H$_2$O in high-$z$ ultra-luminous infrared galaxies}

As in the case of SDP.17b (Omont et al.\ 2011), the strength of the H$_2$O 2$_{02}$--1$_{11}$ and 2$_{11}$--2$_{02}$ 
emission lines in six new high-$z$ ultra-luminous galaxies implies that water vapor has a high abundance in 
these  sources - probably n(H$_2$O)/n(H$_2$) $\gtrsim 10^{-6}$ as in Mrk\,231 (Gonz\'alez-Alfonso et al.\ 2010), 
although this has to be confirmed by detailed modeling - and that its rotation level ladder is excited 
up to energies of at least $\rm \sim 100 \, K$.
This is consistent with the recent results obtained on local ULIRGs such as Arp\,220 and Mrk\,231 
(Rangwala et al.\ 2011; van der Werf et al.\ 2010; van der Werf et al.\ in prep.; Yang et al.\ in prep.). 

Our results 
seem to indicate that high-$z$ ultra-luminous galaxies are analogs of their local counterparts 
scaled by factors up to an order of magnitude in infrared luminosity and star formation rate. 
The similitude in excitation of high-energy H$_2$O levels, observed in local ULIRGs or in APM\,08279+5255 
(van der Werf et al.\ 2011), still needs to be confirmed in high-$z$ sources by future observations of the 
higher frequency H$_2$O lines connected to these levels. 
Nevertheless, the results reported here already indicate many similarities, 
such as  the possibly slightly faster-than-linear increase of the intrinsic  H$_2$O luminosity,  L$_{\rm H_2O}$, with the 
infrared luminosity L$_{\rm IR}$;  the corresponding fact that the ratio L$_{\rm H_2O}$/L$_{\rm IR}$ in high-$z$ sources is 
slightly higher than in local ULIRGs (Fig.\ 3);  and the high intensity ratio of H$_2$O lines to nearby high--J CO lines, 
in between 0.3 to 0.5, as observed in local ULIRGs. 
It is therefore likely that the infrared radiation plays a significant role in the H$_2$O excitation, as demonstrated in the case of Mrk\,231 by the detailed modeling of its spectrum by Gonzalez-Alfonso et al. (2010).

The ubiquity of H$_2$O emission lines in the submillimeter spectra of luminous high-$z$ starburst galaxies 
underlines the importance of water vapor in probing their warm, dense, and dusty interstellar cores and, in
addition, their strong local infrared radiation fields. Indeed, the H$_2$O lines constitute a totally 
different diagnostic from, e.g., CO or low--J HCN emission lines, that are excited by collisions and hence 
trace only temperature and density. 
Because of the large electric dipole of H$_2$O and its large rotational constant, the critical density for collisional excitation of its excited levels is extremely high, typically 10$^8$\,cm$^{-3}$, much higher than expected densities in our sources. The relatively easy detection of H$_2$O emission lines in the sample
of high-$z$ starburst galaxies studied in this paper is therefore indicative of the presence of 
dusty, infrared-opaque nuclear disks in their centers as in local ULIRGs and of a significant role of infrared excitation of H$_2$O (e.g.\ Gonz\'alez-Alfonso et al.\ 2010). 

The role of a strong AGN in the excitation of H$_2$O 
of local ULIRGs is not yet entirely clear 
(van der Werf et al.\ 2010).
There is no definite evidence of the presence of a strong AGN in any of the seven 
sources discussed in this paper, neither from their optical/near-infrared spectra 
nor from their X-ray emission. There is also no indication of a high radio or  mid-infrared excess 
that could have been detected in the FIRST radio survey 
(Becker et al. 1995) or the Wide-field Infrared Survey Explorer (WISE; Wright et al. 2010)  
- see Omont et al.\  (2011) for a detailed discussion of the radio emission of SDP.17b. 
However, in the case of SDP.81, the radio spectral index suggests that a fraction of the radio emission 
is powered by an AGN (Valtchanov et al.\ 2011). 

We stress again that most of our sources are much more powerful than local ULIRGs and that they probe 
a regime of infrared luminosities up to a few 10$^{13}$ L$_{\sun}$, i.e., an order of magnitude higher than 
local ULIRGs. It is probable that the local radiation field is also stronger, and that it might reach the 
limit of stability of such cores against radiation pressure, possibly close to the limit of maximum starbursts. 
As  starbursts with intrinsic far-infrared luminosities much higher than a few 10$^{13}$ L$_{\sun}$ do not appear 
to exist (see e.g.\ Karim et al.\ 2012), it seems that some of the objects that we are studying, such as G15.v2.779, might be close to this 
critical regime.  

All sources studied in this paper are highly magnified by gravitational lensing. This makes the 
detection of lines such as H$_2$O much easier, especially for the highest lensing magnification factors, i.e., the lowest intrinsic 
luminosities. However, for the strongest luminosities, a few 10$^{13}$ L$_{\sun}$, lensing magnification factors go down 
to a few units; e.g.\, $\mu \, \sim \,4$ for G15.v2.779. Therefore, for similarly  luminous sources, 
H$_2$O lines should be easily detectable even in the absence of lensing magnification. For instance, a 
source identical to G15.v2.779 ($\rm L_{\rm IR} \, \sim 2 \times 10^{13} \, L_{\sun}$ at $z>4$) would have $\rm I_{\rm H2O}$\,$\sim$\,1\,Jy.km/s without any magnification, which is well within the reach of the PdBI in its current configuration. 

\section{Conclusion}

This paper reports a significant step forward in the study of water vapor emission in the warm and dense cores of 
ultra-luminous starburst galaxies at high redshift. The seven sources studied here are strongly 
lensed galaxies discovered in the {\it Herschel} H-ATLAS survey and, using the PdBI, all were detected in 
H$_2$O either in the (2$_{02}$--1$_{11}$)  or  the (2$_{11}$--2$_{02}$) rotational line. One of the sources
(SDP.17b) was previously reported in Omont et al. (2011). In all cases, the $\rm H_2O$ emission lines are strong, 
with integrated line fluxes ranging from 1.8 to $\rm 14 \, Jy \, km \, s^{-1}$, comparable both in strength and in 
the characteristics of their profiles to the CO emission lines. The derived apparent infrared  luminosities,
$\rm \mu \, L_{\rm H_2O}$, range from $\sim $3 to $\rm 12 \times 10^8 \, L_{\sun}$. 
When correcting for the magnification factors $\mu$, which are estimated from lensing models based on submillimeter imaging
and corroborated by the empirical relationship between CO luminosity and linewidth, 
the derived H$_2$O luminosities show a strong dependence on the infrared luminosity. It is found that
$\rm L_{\rm H_2O}$ varies as $\sim$L$_{\rm IR}^{1.2}$, the exact shape of the relation remaining uncertain mainly due to 
the uncertainties in the derivation of $\mu$. 
This slightly non-linear relation may indicate that infrared
pumping plays a role in the excitation of H$_2$O lines. Water emission is therefore expected to become very 
strong in the most ultra-luminous (L$_{\rm IR}$\,$>$\,10$^{13}$\,$\rm L_{\sun}$) galaxies at high redshift. 

The results described in this paper underline the fact that the most powerful high-$z$ infrared galaxies are 
analogs of local ULIRGs scaled by factors up to 10 in infrared luminosity and star formation rate. 
The fact that the excitation of H$_2$O levels with E$_{\rm up}$\,$\sim$\,100--150\,K in high-redshift luminous galaxies appears to be similar to what is seen in local ULIRGs is supported by the present results, including the 
rapid increase of  L$_{\rm H_2O}$ with  L$_{\rm IR}$, the slightly higher  
ratio  L$_{\rm H_2O}$/L$_{\rm IR}$ in high-$z$ sources than in local ULIRGs, or the high intensity ratio 
of H$_2$O lines to nearby CO lines. 
However, future observations of the higher frequency H$_2$O lines connected to upper levels in high-$z$ galaxies
are needed to compare the excitation of the high-energy H$_2$O levels with E$_{\rm up}$\,$>$\,150--200\,K  to that of local ULIRGs. 

The detection of $\rm H_2O$ emission lines in seven high-redshift luminous galaxies reported in this paper 
demonstrates that the millimeter/submillimeter lines of water vapor are a new and key 
diagnostic tool to probe both the extreme physical conditions and the infrared radiation in the 
warm dense environments of luminous high-$z$ starburst galaxies.  Follow-up observations of such sources, 
taking advantage of the increased sensitivities that are and will become available with ALMA and the upgraded PdBI (NOEMA) facilities, 
will allow more comprehensive studies of water vapor in high-$z$ luminous galaxies, including unlensed candidates, 
providing essential and new insights into the physical conditions of these galaxies which were present when the universe
was young and have no equivalent in the local universe.

\begin{acknowledgements}
Based on observations carried out with the IRAM Plateau de Bure
interferometer. IRAM is supported by INSU/CNRS (France), MPG (Germany)
and IGN (Spain).  The authors are grateful to the IRAM staff for
their support. US participants in H-ATLAS acknowledge support from NASA 
through a contract from JPL. Italian participants in H-ATLAS acknowledge a 
financial contribution from the agreement ASI-INAF I/009/10/0. 
SPIRE has been developed by a consortium of institutes led
by Cardiff Univ. (UK) and including: Univ. Lethbridge (Canada);
NAOC (China); CEA, LAM (France); IFSI, Univ. Padua (Italy);
IAC (Spain); Stockholm Observatory (Sweden); Imperial College
London, RAL, UCL-MSSL, UKATC, Univ. Sussex (UK); and Caltech,
JPL, NHSC, Univ. Colorado (USA). This development has been
supported by national funding agencies: CSA (Canada); NAOC
(China); CEA, CNES, CNRS (France); ASI (Italy); MCINN (Spain);
SNSB (Sweden); STFC, UKSA (UK); and NASA (USA). M.J.~Micha{\l}owski acknowledges the support of a FWO-Pegasus Marie Curie Fellowship.

\end{acknowledgements}

\end{document}